# Hypotheses for Triton's Plumes:
# New Analyses and Future Remote Sensing Tests

Jason D. Hofgartner*[1], Samuel P. D. Birch[2], Julie Castillo[1], Will M. Grundy[3,4], Candice J. Hansen[5], Alexander G. Hayes[6], Carly J. A. Howett[7], Terry A. Hurford[8], Emily S. Martin[9], Karl L. Mitchell[1], Tom A. Nordheim[1], Michael J. Poston[10], Louise M. Prockter[11], Lynnae C. Quick[8], Paul Schenk[12], Rebecca N. Schindhelm[13], Orkan M. Umurhan[14]

*Corresponding Author (Jason.D.Hofgartner@jpl.nasa.gov)

[1]Jet Propulsion Laboratory, California Institute of Technology, Pasadena CA,
[2]Department of Earth, Atmospheric, and Planetary Sciences, Massachusetts Institute of Technology, Cambridge MA,
[3]Lowell Observatory, Flagstaff AZ,
[4]Northern Arizona University, Flagstaff AZ,
[5]Planetary Science Institute, Tucson AZ,
[6]Department of Astronomy, Cornell University, Ithaca NY,
[7]Southwest Research Institute, Boulder CO,
[8]NASA Goddard Space Flight Center, Greenbelt MD,
[9]Center for Earth and Planetary Studies, National Air and Space Museum, Smithsonian Institution, Washington DC,
[10]Southwest Research Institute, San Antonio TX,
[11]Johns Hopkins University Applied Physics Laboratory, Laurel MD,
[12]Lunar and Planetary Institute, Houston TX,
[13]Ball Aerospace, Boulder CO,
[14]SETI Institute at NASA Ames Research Center, Moffett Field CA,

## Abstract

At least two active plumes were observed on Neptune's moon Triton during the Voyager 2 flyby in 1989. Models for Triton's plumes have previously been grouped into five hypotheses, two of which are primarily atmospheric phenomena and are generally considered unlikely, and three of which include eruptive processes and are plausible. These hypotheses are compared, including new arguments, such as comparisons based on current understanding of Mars, Enceladus, and Pluto. An eruption model based on a solar-powered, solid-state greenhouse effect was previously considered the leading hypothesis for Triton's plumes, in part due to the proximity of the plumes to the subsolar latitude during the Voyager 2 flyby and the distribution of Triton's fans that are putatively deposits from former plumes. The other two eruption hypotheses are powered by internal heat, not solar insolation. Based on new analyses of the ostensible relation between the latitude of the subsolar point on Triton and the geographic locations of the plumes and fans, we argue that neither the locations of the plumes nor fans are strong evidence in favor of the solar-powered hypothesis. We conclude that all three eruption hypotheses should be considered further. Five tests are presented that could be implemented with remote sensing observations from future spacecraft to confidently distinguish among the eruption hypotheses for Triton's plumes. The five tests are based on the: (1) composition and thickness of Triton's southern hemisphere terrains, (2) composition of fan deposits, (3) distribution of active plumes, (4) distribution of fans, and (5) surface temperature at the locations of plumes and/or fans. The tests are independent, but complementary, and implementable with a single flyby mission such as the Trident mission concept. We note that, in the case of the solar-driven hypothesis, the 2030s and 2040s may be the last chance for approximately a century to observe actively erupting plumes on Triton.

## 1. Introduction

Triton, as both a potential Ocean World (OW) and captured Kuiper Belt Object (KBO), is a critical world for comparative planetology. The NASA Roadmap to Ocean Worlds deemed Triton to be the "highest priority target" among candidate OWs "based on the extraordinary hints of activity ... and the potential for ocean-driven activity" (Hendrix et al., 2019). Triton was likely a KBO that was captured by Neptune (McKinnon, 1984; Agnor and Hamilton, 2006). With a radius greater than that of Pluto, Triton is the largest known KBO, and nitrogen-ice on its surface is in vapor pressure equilibrium with its atmosphere, as on Pluto and likely Eris (e.g., Ingersoll et al., 1990; Tegler et al., 2012).

The only spacecraft to explore Triton thus far is Voyager 2 with a single flyby in 1989 (Stone and Miner, 1989). It discovered an enigmatic surface with few impact craters: Triton likely has one of the youngest surfaces in the solar system ($\lesssim$ 10 Myr old; Schenk and Zahnle, 2007). One of the most exciting discoveries from Voyager 2's brief encounter was plumes emanating from Triton's surface, the first active plumes to be discovered on an icy world (Figure 1; e.g., Soderblom et al., 1990). Both plumes had trailing clouds that were > 100 km long and at an altitude of $\approx$8 km. One of the trailing clouds was observed to vary over timescales of less than an hour and to cast a shadow. Nearly a dozen other clouds, with varying degrees of resemblance to the windblown clouds from the plumes, were also observed and possibly sourced from additional plumes (e.g., Hansen et al., 1990). Only two of the clouds were observed to have rising columns from Triton's surface and as such, only two plumes are considered confirmed: they are named Mahilani and Hili and located at 2° E, 49° S and 28° E, 57° S respectively (e.g., Kirk et al., 1995).

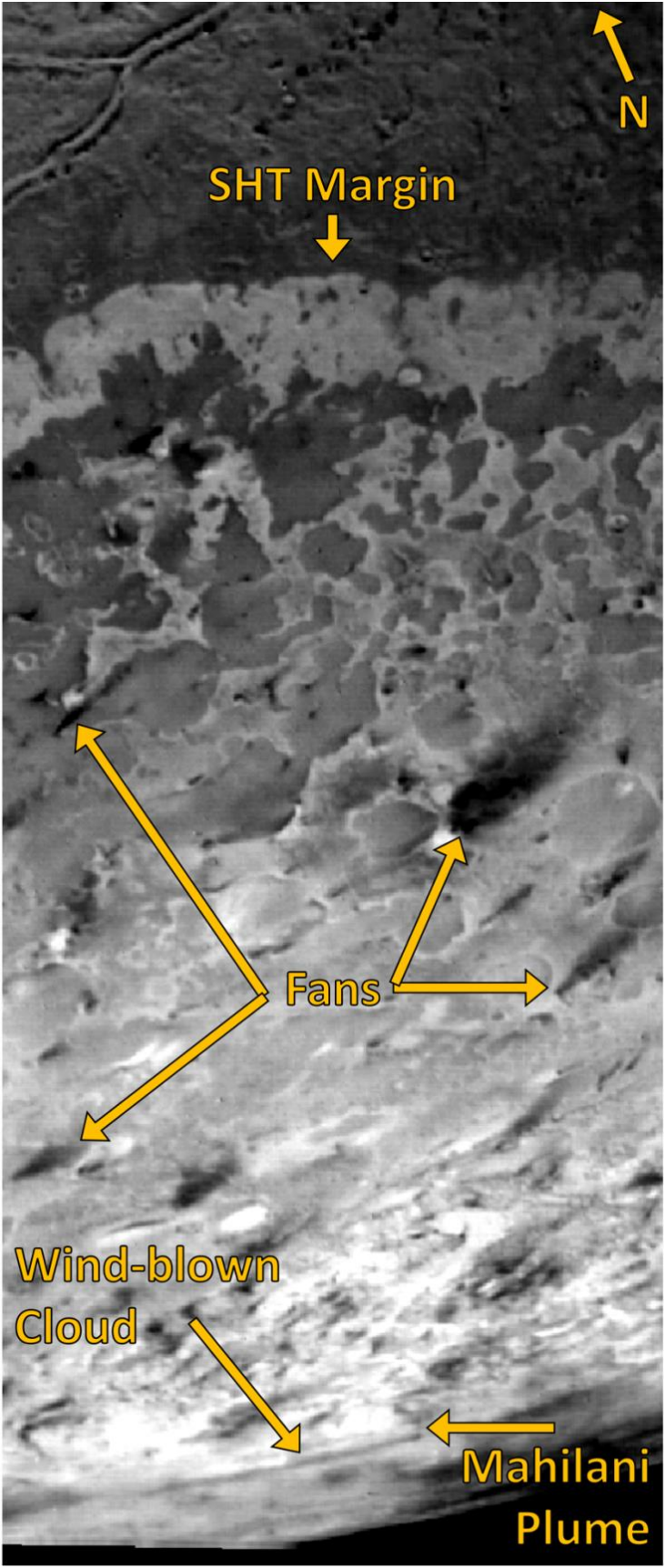

N

SHT Margin

Fans

Wind-blown
Cloud

Mahilani
Plume

*Figure 1: Voyager 2 visible image of part of Neptune's moon Triton in 1989. The image includes the Mahilani plume at 2º E, 49º S and its > 100 km-long trailing cloud, several fans (four of which are annotated but others are not), and the boundary of the high-albedo southern hemisphere terrains (SHT) near the equator. The figure is modified from C1139503 from NASA's Planetary Data System, it has been cleaned, calibrated, and geometrically corrected; no photometric correction was applied.*

Voyager 2 also imaged ≈120 fan-shaped, relatively-dark streaks (called fans and/or dark streaks) on Triton's surface with similar albedo contrasts with their surroundings as the trailing clouds from the plumes (Figure 1; Hansen et al., 1990). These fans varied in length from < 5 km to > 100 km. They are widely interpreted to be deposits from previous plumes (e.g., Hansen et al., 1990; Kirk et al., 1995). If so, they provide crucial additional constraints on the process responsible for the plumes; however, the relationship between fans and plumes is unconfirmed. The two confirmed plume-trailing clouds had a westward direction, whereas the direction of the fans was generally northeast. This difference may be explained by variations in the wind direction with altitude (Ingersoll, 1990). It is also possible that the fans were lower-altitude, active plumes; their altitudes are observationally constrained to < 1 km (Hansen et al., 1990).

In the next section, we briefly introduce several published hypotheses for Triton's plumes and also offer new considerations. In Sections 3 and 4, we scrutinize previously published arguments that the locations of the plumes and fans strongly support a solar-driven mechanism for Triton's plumes and find that the locations are not strong evidence for solar-powered plumes. Section 5 discusses five possible tests, that could be implemented using remote sensing observations by a future mission, to distinguish among current hypotheses for Triton's plumes. In Section 6, we argue that from a Triton-seasonal perspective, the next few decades are a particularly opportune time to explore Triton and apply the tests in Section 5. The final section summarizes the conclusions.

## 2. Hypotheses for Triton's Plumes

There are several published hypotheses for Triton's plumes. Kirk et al. (1995) thoroughly reviewed the models and grouped them into five hypotheses: (1) Insolation-Driven Nitrogen Geysers, (2) Outgassing of the Ice Mantle, (3) Basal Melting of a Permanent Nitrogen Polar Cap, (4) Dust Devils, and (5) Methane Plumes. These hypotheses are briefly introduced in this section and, for the first three, we offer new considerations.

### 2.1. Solar-driven Hypothesis

The geographic distribution of the plumes and fans was argued to be strong evidence for a solar-powered process (e.g., Hansen et al., 1990). The plumes were at latitudes near the subsolar latitude of 45º S during the Voyager 2 encounter. The fans, interpreted to be deposits from former plumes, were generally north of the plumes, broadly distributed over latitudes corresponding to the subsolar latitude in the decades prior to the flyby. The confinement of all plumes and > 100 fans to the high-albedo southern hemisphere terrains (SHT; Figure 1), which were argued to be a nitrogen polar cap (e.g., McEwen, 1990) further suggested an insolation-driven, volatile-dependent process. A solid-state greenhouse model, where a layer of nitrogen-ice is more transparent to incident solar radiation than emitted thermal radiation, which leads to an increase in temperature within and/or at the base of the nitrogen-ice layer, which in turn results in a significant pressure increase due to nitrogen's volatility, which then explosively vents to the atmosphere, emerged as a viable hypothesis (Figure 2; Soderblom et al., 1990; Hansen et al., 1990; Kirk et al., 1990; Brown et al., 1990). This is the most studied model and was previously, for three decades, considered to be the leading hypothesis (e.g., Kirk et al., 1995).

In one variant of this hypothesis, called the "super" greenhouse model by Brown et al. (1990), a dark, absorbing layer underlies a nitrogen-ice layer that is only a few meters thick. In another variant, called the classical greenhouse model by Brown et al. (1990), the nitrogen-ice layer is thicker and translucent such that the greenhouse effect occurs throughout the layer rather than primarily at its base. We group these two variants together and refer to the group as the solar-driven model or hypothesis. Key characteristics of the solar-driven model and two other eruption hypotheses are compared in Table 1.

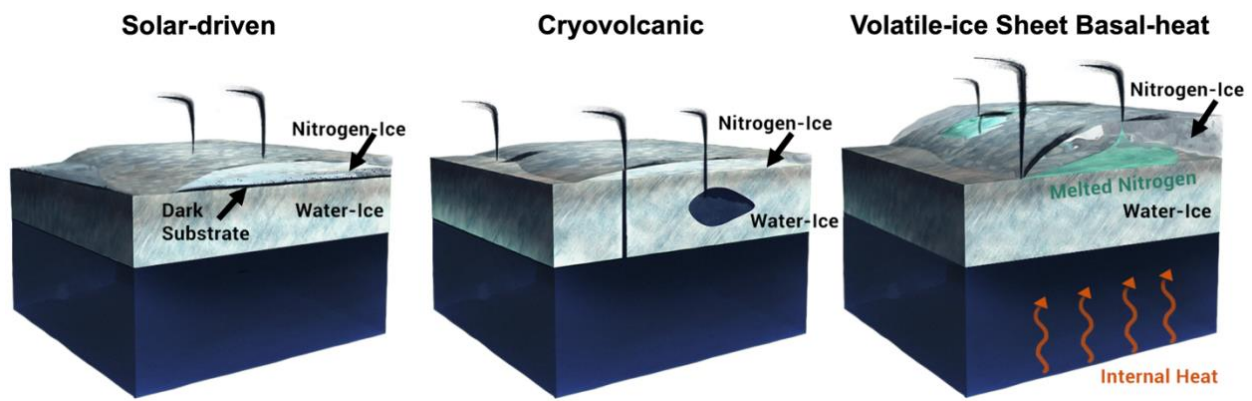

*Figure 2: Artistic depiction of three eruption hypotheses for Triton's plumes. Features are not shown to scale. Nitrogen-ice is assumed to be the most relevant volatile-ice and water-ice is assumed to be the primary refractory material. An ocean is shown beneath the water-ice but is not required. As observed on Triton, the plume-trailing clouds and surface fans point in different directions, which has been explained by differences in wind direction with altitude (e.g., Ingersoll, 1990). This graphic was created by Lizbeth de la Torre and Lisa Poje at the Jet Propulsion Laboratory in consultation with the authors of this paper and is used with permission.*

*Table 1: Key characteristics of three eruption hypotheses for Triton's plumes. Nitrogen is assumed to be the most relevant volatile but substitution by other volatiles such as carbon monoxide, methane, argon, or mixtures of these volatiles may also apply (e.g., Fray and Schmitt, 2009), for all of the hypotheses. Water is assumed to be the primary refractory material but substitution by other refractory materials such as carbon dioxide or mixtures may also apply.*

| Hypothesis | Solar-driven Eruption | Cryovolcanic Eruption | Volatile-ice Sheet Basal-heat Eruption |
|---|---|---|---|
| References | 1 - 5 | 1, 6 | 1, 7, 8 |
| Brief Description | Nitrogen-ice layer causes greenhouse effect, volatility of nitrogen results in significant pressure buildup at base and/or within ice, explosive venting | Explosive outgassing of interior | Internal heat causes solid-state convection and/or melting of nitrogen-ice sheet, rise to surface or flow to margin, explosive sublimation and/or boiling |
| Power Source | Solar insolation | Internal heat | Internal heat |
| Primary Vapor Composition | Nitrogen | Water | Nitrogen |
| Southern Hemisphere Terrains (SHT) Attributes | Nitrogen-ice rich | Formed by or preferentially permit plumes | Nitrogen-ice rich |
| | | Not enhanced in nitrogen-ice | Thick ($\gtrsim$ 100 m) |
| Locations (within SHT for all hypotheses) | Follow seasonal insolation (subsolar latitude) | Independent of seasons | No to weak dependence on seasons |

(1) Kirk et al., 1995; (2) Soderblom et al., 1990; (3) Hansen et al., 1990; (4) Brown et al., 1990; (5) Kirk et al., 1990; (6) Hansen et al., 2021; (7) Brown and Kirk, 1994; (8) Duxbury and Brown, 1997

Like Triton, Mars and Pluto have global, collisional atmospheres in vapor pressure equilibrium with ice on their surfaces (e.g., Hofgartner et al., 2019 and references therein). On Mars, a carbon dioxide analog of the nitrogen, solar-driven model for Triton's plumes generates fans and other seasonal features at both poles (e.g., Kieffer et al., 2006; Hansen et al., 2010; 2013). Plumes and fans similar to those on Triton were predicted to also occur on Pluto's nitrogen-rich surface (e.g., Buratti et al., 2015), but no plumes were observed during the New Horizons flyby (Hofgartner et al., 2018). However, the subsolar latitude on Pluto during the New Horizons flyby in 2015 was 52º N and moving northward, while the nitrogen-rich Sputnik Planitia (Grundy et al., 2016) extends from ≈22º S to 50º N. Thus, one possible explanation for the observation of plumes on Triton but not Pluto is that the seasonal timing was suitable for observing plumes with Voyager 2 at Triton but not for New Horizons at Pluto. Darks streaks with some resemblance to those on Triton were observed on Pluto, including Sputnik Planitia, but they are generally smaller in scale and less numerous; some are also associated with topographic wind obstacles, but no such association is apparent for others (Hofgartner et al., 2018). Thus, of the three worlds in the solar system with a global, collisional atmosphere in vapor pressure equilibrium with ice on the surface: Mars has fans that result from the solid-state greenhouse effect, Triton has plumes and fans that are hypothesized to be products of that effect, and possible evidence that it occurs on Pluto is much weaker.

## 2.2. Cryovolcanic Hypothesis

A second eruption hypothesis for Triton's plumes is that they are a manifestation of explosive cryovolcanism (Figure 2; Table 1; Kirk et al., 1995). The occurrence of cryovolcanism on an icy world had not been definitively observed or confirmed at the time of the Voyager 2 flyby of Triton in 1989, but today, Enceladus is known to have a cryovolcanic plume (e.g., Porco et al., 2006). Moreover, explosive and/or effusive cryovolcanism is a leading hypothesis for numerous features on Triton (e.g., Croft et al., 1995) and throughout the outer solar system (e.g., Geissler, 2015 and references therein). In addition to direct observation of a cryovolcanic plume from Enceladus, two more recent arguments in favor of the cryovolcanic hypothesis for Triton's plumes are: their mass flux, and growing evidence that Triton is an ocean world (Hendrix et al., 2019; Hansen et al., 2021). The estimated mass flux of Triton's plumes (possibly exceeding 400 kg/s) is much greater than that of Mars' solar-driven analogs ($\approx$0.2 kg/s) but similar to that of Enceladus' cryovolcanic plume ($\approx$200 kg/s; Hansen et al., 2021 and references therein). Obliquity tides that result from Triton's inclined orbit could be a significant source of internal energy, different from radiogenic or eccentricity tides on other natural satellites, that may maintain a subsurface ocean and power cryovolcanism (Nimmo and Spencer, 2015).

Advances in knowledge of cryovolcanism, as well as Triton's interior structure and heating, warrant new consideration of the cryovolcanic hypothesis for Triton's plumes. However, we note that for this hypothesis: (1) the fans are not from plumes and/or (2) the SHT is not expected to be enriched in nitrogen-ice (or any other seasonally volatile-ice such as carbon monoxide or methane). If Triton's fans are from former plumes, as is widely taken to be the case (e.g., Kirk et al., 1995; Croft et al., 1995), then the plumes are very likely related to the SHT. Recall that $\approx$120 fans were identified (Hansen et al., 1990) and all, without exception, were on the SHT; the fans are also distributed widely across the SHT (the SHT extent and fan geographic distribution are shown and discussed further in Section 4). The fans are spectrally (McEwen, 1990) and also photometrically (Hillier et al., 1994) distinct from other terrain units on Triton, so they would be detectable outside of the SHT had they existed in other regions observed by Voyager 2. The broad distribution of the fans yet perfect confinement to the SHT is not a coincidence[1]. Thus, if Triton's plumes are cryovolcanic and the fans are produced by plumes, then the SHT must be related to cryovolcanism in some way: cryovolcanism formed the SHT and/or the SHT preferentially permits plume/fan forming cryovolcanism. Since the cryovolcanic hypothesis does not require nitrogen-ice, but the plumes/fans are clearly confined to the SHT, we consider the cryovolcanic hypothesis to effectively predict the SHT is not enhanced in nitrogen-ice (or other volatile-ice). The composition of the SHT was not directly measured by Voyager 2 because its infrared spectrometer was not capable (due to wavelength coverage and sensitivity) of constraining the composition of features identified on Triton.

## 2.3. Volatile-ice Sheet Basal-heat Hypothesis

A third hypothesis for Triton's plumes also posits that internal energy drives eruptions, as in the cryovolcanic hypothesis, but is critically dependent on nitrogen-ice. In the basal heating of a nitrogen-ice sheet hypothesis, Triton's internal energy heats the base of a thick ($\approx$ km) layer of nitrogen-ice and plumes result when nitrogen that is warmer than the atmosphere is transported to the surface (Figure 2; Table 1). In one variant of this hypothesis, the nitrogen and heat are transported in the liquid phase: the base of the nitrogen-ice layer melts and the melt is exposed to the atmosphere after buoyant vertical rise through the nitrogen-ice and/or flow at the base to the margin (Brown and Kirk, 1994; Kirk et al., 1995). In another variant, the nitrogen and heat are transported in the solid phase via solid-state convection (Duxbury and Brown, 1997). We group these two variants together and refer to the group as the volatile-ice sheet basal-heat model or hypothesis to emphasize that basal-heat (internal heat) is the primary power source and basal here refers to the base of a thick volatile-ice layer, not the lithosphere.

The volatile-ice sheet basal-heat hypothesis is enticing because of its association with nitrogen-ice, which is prevalent on Triton and argued to be abundant in the SHT (e.g., McEwen, 1990), and because the eruptions are powered by internal heat, which is predicted to be significant as a result of obliquity tides (Nimmo and Spencer, 2015). However, ongoing solid-state convection is likely resurfacing Pluto's nitrogen-rich ice sheet, Sputnik Planitia (McKinnon et al., 2016), and yet it does not have plumes/fans analogous to those on Triton (Hofgartner et al., 2018). The lack of analogous plumes/fans on Sputnik Planitia may be evidence against the solid-state convection variant of this hypothesis for Triton's plumes, but more work comparing Triton and Pluto is needed before drawing such a conclusion. For example, Triton's greater internal heat flow compared to that of Pluto (e.g., Nimmo and Spencer, 2015) may result in more vigorous convection (and/or greater melting), which may drive explosive eruptions. Triton's greater internal heat flow may also explain its plumes/fans differences with Pluto in the context of the cryovolcanic hypothesis. Brown and Kirk (1994) and Duxbury and Brown (1997) specifically studied a pure nitrogen-ice layer but the model may be generalizable to other volatile-ices on Triton such as carbon monoxide, methane, argon, or a mixture (i.e., ices that are volatile on Triton and can be transported by the atmosphere in the vapor phase, not refractory ices on Triton such as water and carbon dioxide).

## 2.4. Dust Devil Hypothesis

The dust devil model hypothesizes that Triton's plumes are atmospheric vortices with entrained dust, not eruptions (Ingersoll and Tryka, 1990). This model utilized a preliminary temperature profile of the atmosphere that was subsequently demonstrated to be incorrect (Kirk et al., 1995). The model has also been criticized because the winds in the atmospheric vortices are not strong enough to pick up dust (i.e., the dark material in plumes and fans) and to prevent shear of the vertical column, which was not observed (Kirk et al., 1995). Thus, we do not consider this hypothesis further.

## 2.5. Buoyant Methane Hypothesis

A second atmospheric, not eruptive, hypothesis for Triton's plumes is buoyant methane that sublimated from the surface. This atmospheric hypothesis faces similar criticisms as the dust devil hypothesis: that winds are neither strong enough to pick up dust nor prevent vertical shear (Kirk et al., 1995). The eruptive hypotheses allow for significantly stronger plume winds because plume material can be pressurized beneath the surface before contact with the atmosphere, unlike the atmospheric hypotheses, enabling greater initial pressures and thus winds. The buoyant methane model is briefly reviewed in Kirk et al. (1995) but their primary reference was in press and not subsequently published, and it appears the model has not been published in other peer-reviewed literature. We do not consider it further.

## 3. Latitude of Solar-driven Plumes

The proximity of Triton's plumes to the subsolar latitude at the time of the Voyager 2 encounter was considered a key argument in favor of the solar-driven hypothesis (e.g., Kirk et al., 1990; Kirk et al., 1995). In this section, we scrutinize that evidence and argue that the similarity of the latitude of the plumes to the subsolar latitude is not strong evidence in favor of the solar-driven model.

The maximum instantaneous solar insolation on a sphere is at its subsolar point; however, the maximum diurnal insolation is not always at the subsolar latitude. Figure 3A shows the diurnal-average insolation on a sphere at all latitudes, for various subsolar latitudes. When the subsolar latitude is 0°, both the maximum instantaneous insolation and maximum diurnal-average insolation are at the equator. As the subsolar latitude moves away from the equator, the latitude of the maximum diurnal-average insolation moves in the same direction but no longer coincides with the subsolar latitude; it is further from the equator. The latitudes of maximum instantaneous insolation and maximum diurnal-average insolation are similar until the diurnal-average jumps to the pole at a subsolar latitude of 20.7°. For all subsolar latitudes poleward of 20.7°, the illuminated pole receives the maximum diurnal-average insolation. This fact is somewhat unfamiliar to terrestrial experiences where the subsolar latitude only ranges by +/- 23.5° (and the difference in diurnal insolation between the pole and subsolar latitude is mild between 20.7° and 23.5°) and because terrestrial weather is affected by many other factors. However, it is important for many worlds in the outer solar system that have much larger obliquities (the difference between the illuminated pole and subsolar latitude approaches 50% near a subsolar latitude of 40°). Thus, the diurnally-averaged flux does not predict that solar-driven plumes would be near the subsolar latitude of 45° S during the Voyager 2 encounter, but rather near the south pole. The confirmed plumes, Mahilani and Hili at 49° S and 57° S, are much closer to the subsolar latitude than they are to the south pole, inconsistent with expectations based on maximum diurnal solar flux.

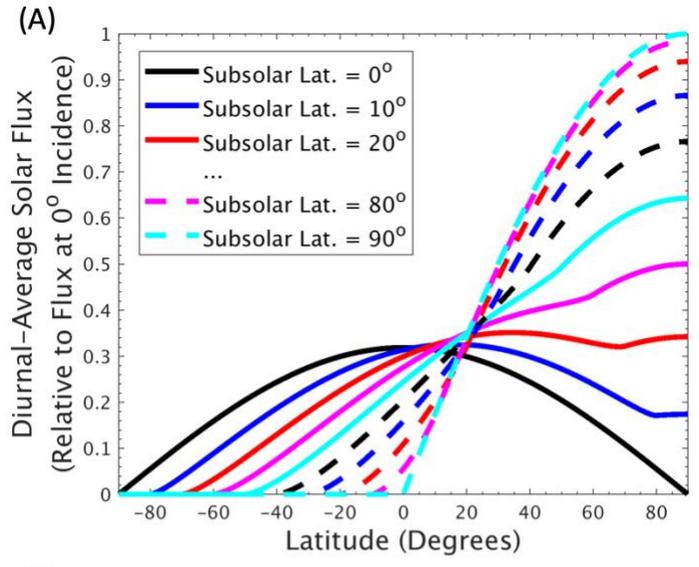

(A)

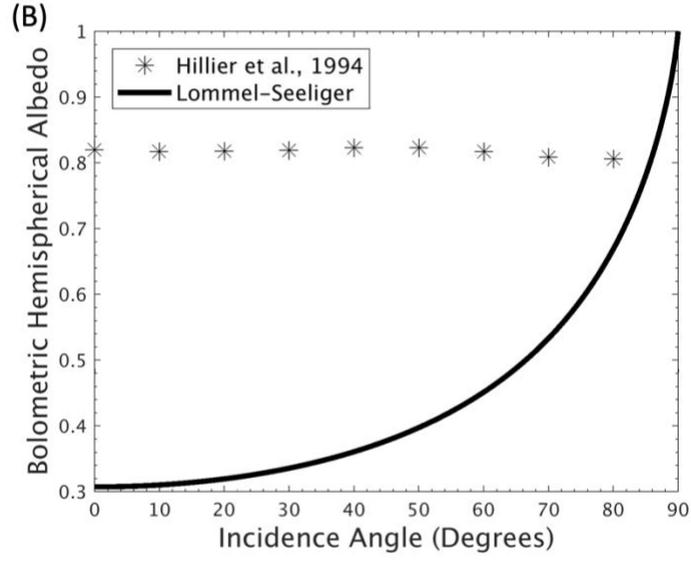

(B)

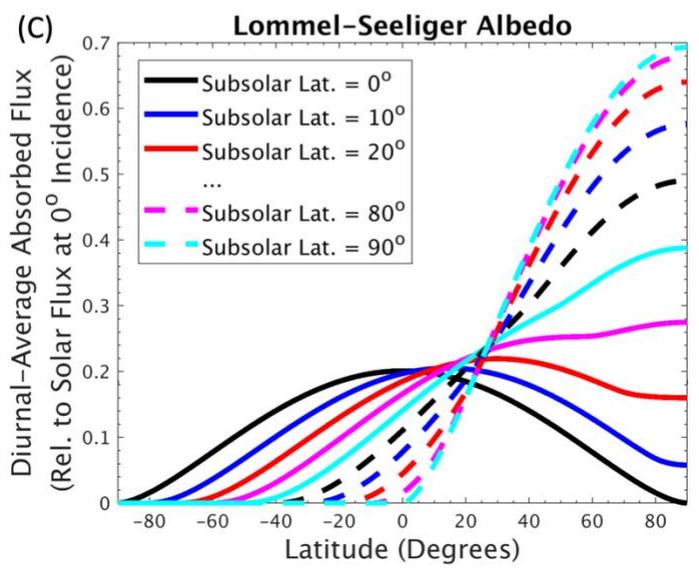

(C)

*Figure 3: Subsolar latitude and latitude of maximum diurnal-average flux. (A) Diurnal-average solar flux as a function of latitude and subsolar latitude. Subsolar latitudes of 0ºN, 10ºN, 20ºN, 30ºN, 40ºN, 50ºN, 60ºN, 70ºN, 80ºN, and 90ºN are shown. (B) Bolometric (energy balance) albedo for Triton as reported in Hillier et al. (1994) and a Lommel-Seeliger surface. (C) Diurnal-average solar flux absorbed, for the Lommel-Seeliger bolometric albedo in (B). Note the change in scale of the ordinate axis from (A). The latitude at which the absorbed solar flux, averaged over a full rotation, is greatest, is not always near the subsolar latitude.*

The concern of the latitude of maximum diurnal-average insolation was noted in Brown et al. (1990) for the "super" greenhouse model (the variant of the solar-driven hypothesis where a dark, absorbing layer underlies a nitrogen-ice layer that is only a few meters thick; Section 2.1). The classical greenhouse model in Brown et al. (1990) (the solar-driven hypothesis with a thicker, translucent nitrogen-ice layer; Section 2.1) can predict a preference near the Voyager 2 subsolar latitude but we note that this preference is weak. That model used a very low bolometric albedo (35%, whereas more recent estimates are > 80%; e.g., Hillier et al., 1994; Buratti et al., 2011), resulting in liquid nitrogen at depth. The liquid is ≈20 m deep at the subsolar latitude and ≈30-40 m deep elsewhere; it is unclear why liquid at 20 m, but not 30-40 m, would result in eruptions. Kirk et al. (1995) also noted this concern: that the classical greenhouse model requires tuning of the model parameters to explain the observation of plumes only in the mid-southern latitudes. The similar latitude of Triton's plumes and the subsolar latitude during the Voyager 2 encounter, argued to be strong evidence for the solar-driven model, may require more detailed thermal physics to explain, or it may be a coincidence.

The bolometric (energy balance) albedo is an important parameter for the temperature of a planetary surface. This quantity is often taken to be a constant but, except in special cases, does vary with the incidence angle of the insolation (e.g., Squyres and Veverka, 1982). It is conceivable that this consideration could move the latitude of maximum diurnal-average insolation from the south pole toward the subsolar latitude of 45° S during the Voyager 2 encounter. The incidence angle dependence of bolometric albedo for Triton's terrains is reported in Hillier et al. (1994) based on a radiative transfer model fit to Voyager 2 observations: Triton's global-average dependence is shown in Figure 3B. That modeled dependence is too weak to significantly affect the latitude dependence in Figure 3A. We also considered strong variations, exceeding a factor of 3 over the full incidence angle range, and a variety of functional forms, but found that the latitude of maximum diurnal-average insolation does not change appreciably. The albedo dependence and resultant diurnal-average absorbed flux for one example are shown in Figures 3B and 3C respectively. We could create functions that move the maximum from the south pole to the latitudes of the observed plumes for the Voyager 2 encounter subsolar latitude of 45° S but only by fine-tuning the incidence angle dependence of the albedo with excessively strong functions and/or excessively large average albedos, neither of which have a plausible basis in reality. We thus conclude that the incidence angle dependence of bolometric albedo does not have a significant effect on the latitudinal dependence of the diurnal-average insolation shown in Figure 3A. Other thermal considerations may explain the apparent proximity of the subsolar latitude and latitude of the plumes on Triton during the Voyager 2 flyby, or it may be a coincidence.

## 4. Geographic Distribution of Triton's Fans

In addition to the latitude of Triton's plumes, the geographic distribution of its fans was considered strong evidence in favor of the solar-driven hypothesis (e.g., Hansen et al., 1990; Kirk et al., 1995). Below, we scrutinize that evidence and argue that it does not strongly favor the solar-driven model.

The distribution of Triton's fans was mapped by Hansen et al. (1990) and we use their data in this analysis; the distribution is shown in Figure 4A. The fans were mapped in high spatial resolution images of the Voyager 2 flyby-encounter hemisphere (approximately the sub-Neptune hemisphere) of Triton. Fans may have also existed elsewhere on Triton, such as other longitudes of the SHT and closer to the south pole, but Voyager 2 images had insufficient spatial resolution for mapping of fans outside of the flyby-encounter hemisphere (Hansen et al., 1990). The boundary of the SHT was mapped in McEwen (1990) and we use that boundary in this analysis; the boundary is also shown in Figure 4A. Improvements in the Voyager 2 and Triton ephemeris have resulted in shifts from the mapping in Hansen et al. (1990) but since these are approximately independent of location and we are primarily interested in the relative distribution, we ignore this inaccuracy of the absolute locations. Similarly, a possible absolute shift between the Hansen et al. (1990) and McEwen (1990) datasets is ignored.

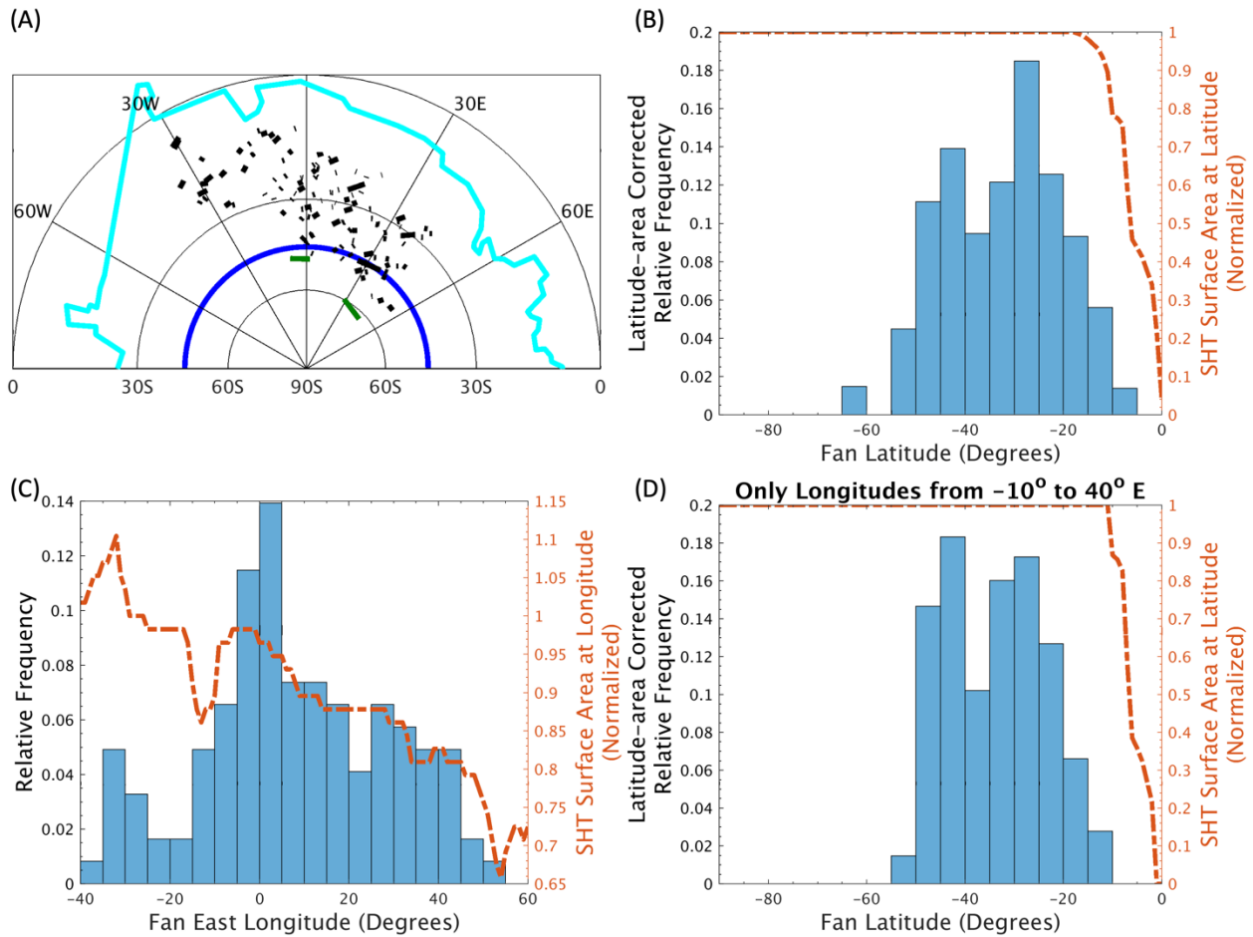

*Figure 4: Geographic distribution of Triton's fans. (A) Triton's fans as mapped in Hansen et al. (1990) are shown in a stereographic projection centered at the south pole. Fans are shown in black and the Mahilani and Hili plumes in green. The boundary of the southern hemisphere terrains (SHT) as mapped in McEwen (1990) is shown in cyan. The subsolar latitude of 45° S during the Voyager 2 encounter in 1989 is shown in blue. (B) Histogram of fans and plumes as a function of latitude. The fraction of longitudes (areal fraction), at a given latitude, that are included in the SHT is also shown; only longitudes between the minimum (-35° E) and maximum (53° E) of the mapped fan distribution were included in the fractional coverage. A similar figure but limited to longitudes from -10° to 40° E for both fans/plumes and SHT coverage is shown in (D). (C) Histogram of fans and plumes as a function of longitude. The fractional area, at a given longitude, that is part of the SHT is also shown; a normalized area of unity corresponds to SHT coverage from the south pole to the equator. The imaging quality degrades toward the southwest (e.g., Schenk et al., 2021), so the absence of fans poleward of ≈50° S in (B) and (D) and loss of correlation between fan abundance and SHT coverage west of ≈15° E in (C) are consistent with observational bias. The observed, geographic distribution of Triton's fans may be controlled by the extent of the SHT and observational bias.*

Figure 4B includes a histogram of the relative frequency of fans and plumes as a function of latitude. The total surface area of each latitude bin on a sphere varies from a maximum at the equator to a minimum at the poles and thus more equatorial latitudes on Triton have greater surface area for fans. The histogram relative frequencies are corrected, assuming Triton is a sphere, to account for this changing area with latitude. The latitudes of the apexes of the fan-shaped, relatively-dark streaks, assumed to be the locations of the eruptions that formed the fans, are used. The fans are small enough that the distributions in Figure 4 are not sensitive to this choice to use the apexes. Figure 4B shows that the fans are widely distributed in latitude across the SHT with a peak around 30º S. The imaging quality degrades toward the south pole, so the absence of fans poleward of ≈50º S may be an observational bias. The SHT surface area at each latitude, expressed as a fraction of the longitudinal coverage, is also shown in Figure 4B. The similar change of the fan histogram and SHT curve from the equator to ≈20º S suggests that SHT coverage controls the abundance of fans at these latitudes. The ≈5º space between the histogram and curve could indicate that (detectable) fans are less common near the SHT boundary but may also be an artifact of any shifts between the Hansen et al. (1990) and McEwen (1990) coordinates.

The geometry of the Voyager 2 flyby of Triton constrained images of the surface to vary significantly in spatial resolution, somewhat as a function of longitude (e.g., Schenk et al., 2021). To check that incomplete latitudinal coverage, at different longitudes, did not significantly bias the histogram in Figure 4B, we produced similar figures for more restrictive ranges in longitude. Figure 4D is one example. Generally, the trends in both Figures 4B and 4D of increasing fan density from the equator to a peak at ≈25-30º S, then decreasing to a minimum at ≈35-40º S, followed by increasing density, and then gradual decrease toward the south pole was consistent. This suggests that the distribution shown in Figure 4 is representative of the flyby-encounter hemisphere, from north of the equator to ≈50º S, and not strongly biased by the observations. The histograms may, however, be biased by the sub-Neptune hemisphere as a whole and longitudes ≳ 90º from this hemisphere may have a different fan distribution.

Figure 4C includes a histogram of the fans and plumes as a function of longitude. The latitude-weighted areal coverage of the SHT at each longitude is also shown. The similarity between the histogram and curve between ≈ -15 to 55º E strongly suggests that the SHT area influences and possibly controls the distribution of Triton's fans. The loss of correlation between fan abundance and SHT coverage further west is consistent with an observational bias due to gradually worsening spatial resolution at the southwestern extent of the mapping, whereas the eastern boundary of the high spatial resolution images has a more abrupt edge (e.g., Schenk et al., 2021).

We conclude from this analysis that Triton's fans are indeed broadly distributed across its SHT and that the mapped distribution is also controlled by the SHT boundary and observational biases (e.g., spatial resolution), as noted in previous work. We emphasize, however, the influence of observational bias and interpret it to considerably weaken the evidence, based on the geographic distribution of the fans and plumes, for the solar-driven hypothesis. The inference that the fans were at latitudes consistent with the subsolar latitude in the decades prior to the Voyager 2 encounter, may instead be attributed entirely to a broad fan distribution over the sub-Neptune SHT and observational bias of the Voyager 2 imaging coverage. The distribution of the SHT itself may be controlled by solar energy but the fans may distribute across the SHT independent of the subsolar latitude.

**5. Future Remote Sensing Tests of Hypotheses for Triton's Plumes**

In the preceding sections, we noted two significant concerns with key arguments for favoring the solar-driven hypothesis. The process responsible for Triton's plumes is yet to be confidently determined. In this section, we argue that the three eruption hypotheses for Triton's plumes in Table 1 can likely be confidently distinguished with remote sensing from another single-flyby mission. We emphasize that the three hypotheses in Table 1, with different power sources and roles of volatiles can be distinguished with a flyby; detailed investigations of the eruption mechanisms and distinguishing between slight model variations of the hypotheses may not be possible with a single flyby. We discuss five tests of the predictions of the models that could be implemented by a future flyby (or multi-flyby, or orbiter) mission with suitable remote sensing instrumentation. The five tests are independent, but complementary, which increases resiliency in the event of unexpected complications. Predictions for each of the hypotheses for the five tests are summarized in Table 2.

*Table 2: Testable predictions, with remote sensing, of hypotheses for Triton's plumes. Fans are assumed to be deposits from plumes. Nitrogen is assumed to be the most relevant volatile but substitution by other volatiles such as carbon monoxide, methane, argon, or mixtures of these volatiles may also apply, for all of the hypotheses. Water is assumed to be the primary refractory material but substitution by other refractory materials such as carbon dioxide or mixtures may also apply. Tholin is used to represent dark, complex organic material.*

| Hypothesis | | Solar-driven | Cryovolcanic | Volatile-ice Sheet Basal-heat |
|---|---|---|---|---|
| Hypothesis Tests & Predictions | Composition & Thickness of Southern Hemisphere Terrains (SHT) | Nitrogen-ice rich | Not enhanced in nitrogen-ice | Nitrogen-ice rich |
| | | | | Thick ($\gtrsim$ 100 m) |
| | Composition of Fans | Tholin rich &/or nitrogen-ice rich | Water-ice rich | May include tholins, volatile-ices, & underlying refractory materials (e.g., water-ice) |
| | | | Water-ice may be amorphous | |
| | | | May include water solutes | |
| | Distribution of Active Plumes | Confined to latitude band consistent with seasonal insolation | Independent of seasons | Independent of seasons |
| | | Only from nitrogen-ice rich locations | Independent of nitrogen-ice distribution | Only from nitrogen-ice rich locations |
| | Distribution of Fans | If SHT persists, ubiquitous new fans on SHT, at latitudes south of ≈ subsolar latitude | No significant change | No significant change |
| | | | Independent of nitrogen-ice distribution | Only from nitrogen-ice rich locations |
| | Anomalous Temperatures | Temperature $\lesssim$ 104 K | Temperature can exceed 104 K | Temperature $\lesssim$ 104 K |

## 5.1. Composition and Thickness of Southern Hemisphere Terrains Hypothesis Test

One test of the hypotheses for Triton's plumes is the composition and thickness of the SHT (Figure 5). This test assumes the fans are deposits from plumes, as is widely taken to be the case (e.g., Kirk et al., 1995), and is based on the observation that all fans (> 100; Hansen et al., 1990) and plumes are confined to the SHT, which is highly unlikely to be a coincidence (Sections 2.2 and 4). The solar-driven model is consistent with an SHT that is a thin (< 10 m) seasonal layer or thick (non-seasonal) layer of nitrogen-ice, whereas the volatile-ice sheet basal-heat model requires a thick (≈ km) nitrogen-ice layer (e.g., Kirk et al., 1995). The cryovolcanic model does not predict any difference between the volatile abundance of the SHT and terrains that do not have plumes and fans (e.g., Kirk et al., 1995). Thus, significant retreat of the SHT due to the current southern summer is inconsistent with the cryovolcanic and volatile-ice sheet basal-heat hypotheses. Such a change in SHT extent could be identified even with images at very coarse (e.g., > 100 km) spatial resolution, which is achievable with, in addition to a future spacecraft mission, $\gtrsim$ 20 m aperture ground-based telescopes. If, however, the SHT is predominantly unchanged since the Voyager 2 flyby, then spatially resolved measurement of the composition of the SHT and other terrains can distinguish the cryovolcanic model from the other two. A significant enhancement of volatiles rules out the cryovolcanic model, whereas no relative enhancement or a depletion of volatiles rules out the other two models. Again, the large extent of the SHT allows for this compositional measurement at very coarse (e.g., > 100 km) spatial resolution.

If the cryovolcanic model is ruled out, the solar-driven and volatile-ice sheet basal-heat models may be distinguishable by the thickness of the SHT. A thin layer rules out the volatile-ice sheet basal-heat model but a thick, volatile-rich layer may be ambiguous between these two models. A value of 100 m is used to distinguish thick and thin layers in Figure 5 but it is approximate, as the thickness required for the volatile-ice sheet basal-heat model depends, among other parameters, on the heat flow from the interior and volumetric composition of the SHT and is not presently strongly constrained (Brown and Kirk, 1994; Duxbury and Brown, 1997).

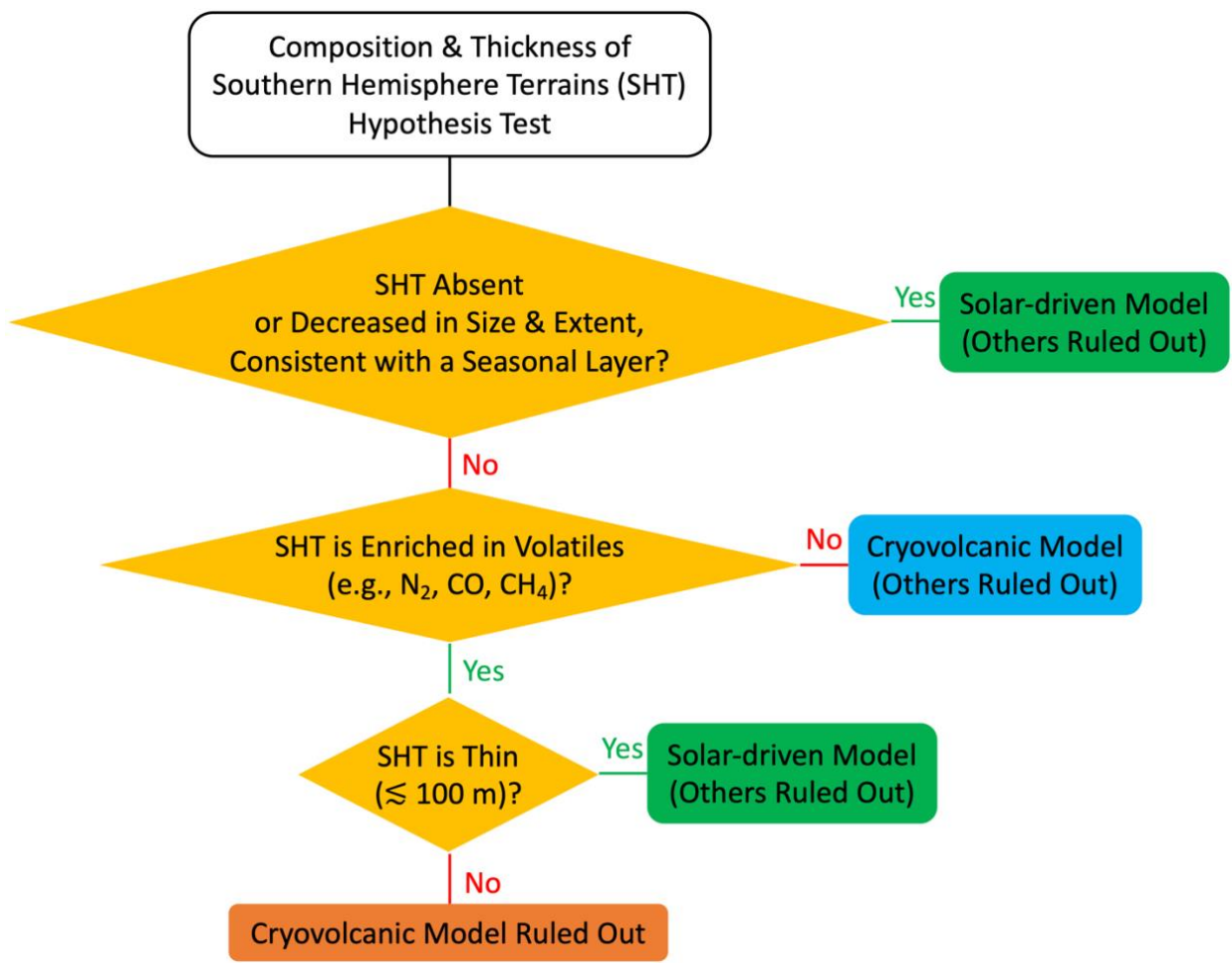

*Figure 5: Composition and thickness of southern hemisphere terrains (SHT) hypothesis test. This test assumes the fans are deposits from plumes.*

5.2. Composition of Fans Hypothesis Test

Assuming the fans are deposits from plumes (e.g., Kirk et al., 1995), their composition is a second test of the hypotheses (Figure 6). The cryovolcanic model predicts fans that are rich in water-ice. Since a cryovolcanic plume would include water vapor that rapidly condenses in the cold Triton atmosphere, its fan deposit may include amorphous water-ice, as observed at Enceladus (Newman et al., 2008). If the water source of a cryovolcanic plume includes dissolved solutes, they may also be present in the fans. This composition test can be applied using near-infrared spectral observations at approximately the scale of Triton's fans (average length of 26 km, some are > 100 km in length; Hansen et al., 1990).

If no fans are identified (e.g., fans observed by Voyager 2 are no longer present), this test is neutral. The predicted eruption frequencies and fan lifetimes for the different hypotheses could be compared to determine whether this null result favors one hypothesis, but since this is less definitive and depends on modeling assumptions, we do not pursue it further. The test is also neutral if the water-ice abundance of a fan is similar to its surroundings. In this case, confounding factors such as sublimation/condensation and aeolian processing may influence the interpretation but may not be well constrained.

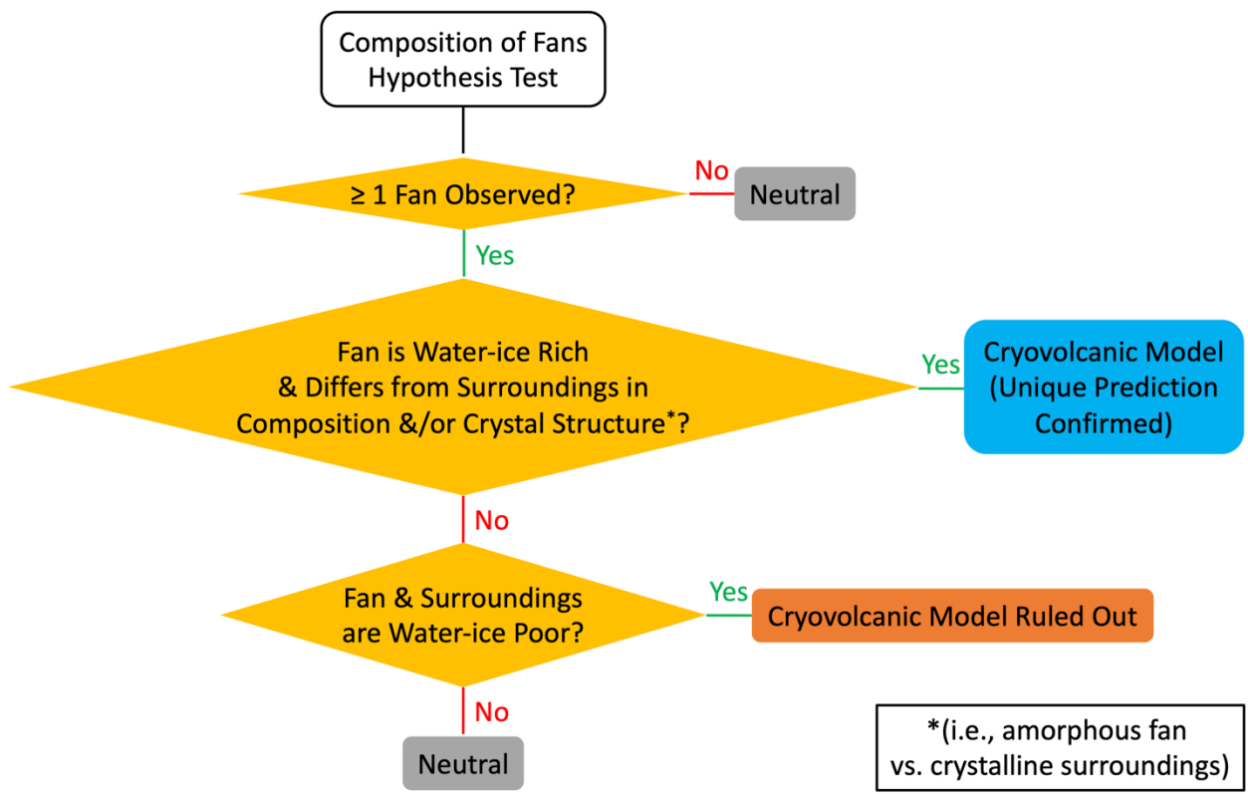

*Figure 6: Composition of fans hypothesis test. This test assumes the fans are deposits from plumes.*

The SHT and fan composition hypothesis tests are both founded on the question of volatile (nitrogen, but possibly others such as carbon monoxide, methane, argon, and mixtures) vs.

refractory (water, but possibly others such as carbon dioxide as well as mixtures) abundance, of the SHT and fans respectively. This relatively rudimentary basis for these hypothesis tests reflects the current poor knowledge of the composition of Triton's surface at resolved spatial scales. The Voyager 2 infrared spectrometer was not sufficiently capable to measure the composition of Triton's surface, so compositional knowledge is limited to unresolved and barely-resolved (few pixels) Earth-based telescopic observations (e.g., Holler et al., 2016) and indirect techniques (e.g., McEwen, 1990). Thus, even a single flyby with a suitably capable near-infrared spectrometer offers a transformative increase in the understanding of the surface composition of Triton's enigmatic terrains, with significant implications for Triton's plumes.

## 5.3. Distribution of Active Plumes Hypothesis Test

The distribution of any active plumes enables a third test of predictions of the hypotheses (Figure 7). The solar-driven and volatile-ice sheet basal-heat hypotheses are critically dependent on nitrogen-ice (or other volatile-ice); therefore, a plume erupting from a location without nitrogen-ice rules out those two models. The solar-driven hypothesis predicts a migration of Triton's plumes that follows seasonal changes of solar insolation. Eruptions could change location in the cryovolcanic and volatile-ice sheet basal-heat models, but since they are powered by internal heat, they would not follow the seasonal solar insolation and entirely new eruptions sites would only be expected on geologic timescales (i.e., greater than the 165 Earth-year orbital/seasonal timescales). The subsolar latitude on Triton was 45° S and moving southward during the 1989 Voyager 2 flyby, while in 2021 it is 36° S and moving northward; any mission to Triton in this century will be at a different season than that of Voyager 2 (see Section 6 for further discussion). Thus, plumes confined to a latitude band consistent with the seasonal insolation, away from the sites of the Voyager 2 plumes and fans, would be compelling evidence for the solar-driven model. Plumes anywhere else on Triton, including sustained eruptions, as observed at Enceladus (e.g., Kite and Rubin, 2016), at the locations of the Mahilani and Hili Voyager 2 plumes, rules out the solar-driven model.

Just as for the fan composition hypothesis test, there is a neutral result of zero active plumes, though arguments based on likelihood (e.g., comparison of eruption frequencies and fan lifetimes) may provide partial constraints. The active plumes could be identified using images; the Voyager 2 active plumes are apparent in stereo images, limb images, temporally separated images, and also produced a shadow and these detection methods are also expected to apply for future investigations.

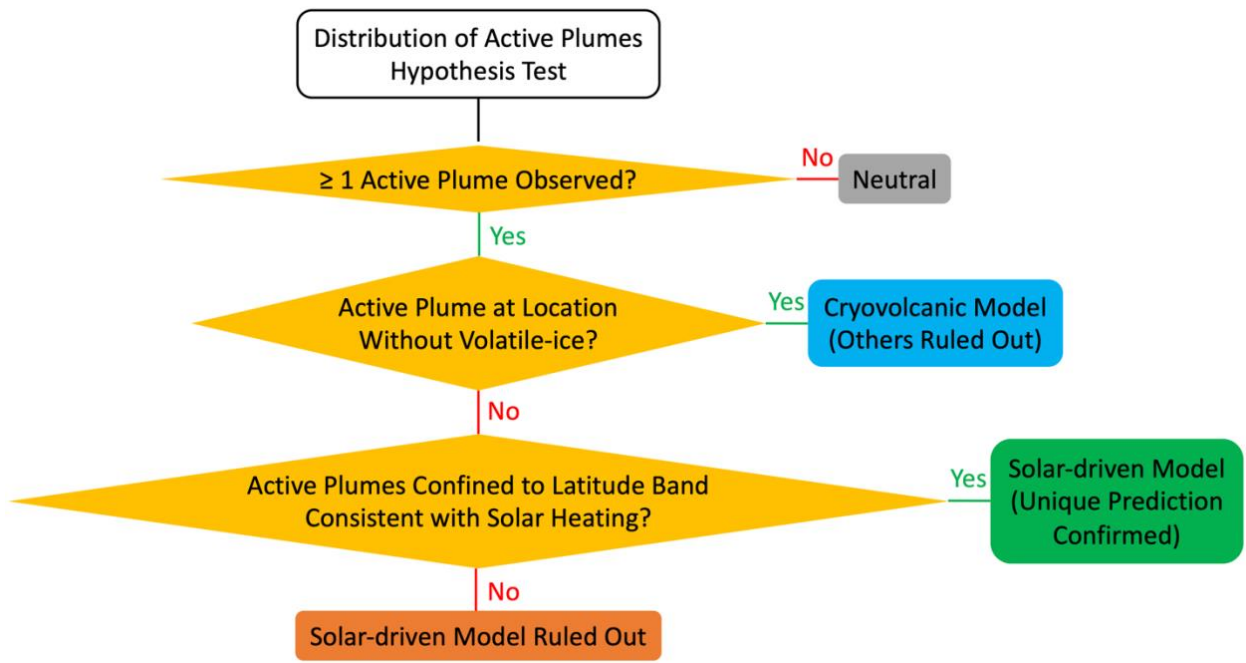

*Figure 7: Distribution of active plumes hypothesis test.*

## 5.4. Distribution of Fans Hypothesis Test

Following similar arguments as for active plumes, the distribution of fans is another test of the hypotheses for Triton's plumes (Figure 8). Like the other tests based on Triton's fans, this test assumes the fans are plume deposits. After the Voyager 2 encounter in 1989, the subsolar latitude moved from 45° S to 50° S in 2001 and has since been moving northward. The next equinox will be in 2046 (discussed further in Section 6). Thus, by the time another spacecraft approaches Triton, the change in subsolar latitude since the Voyager 2 flyby will have included a traverse over many (possibly all) latitudes of the SHT. The changing insolation may have driven a plethora of new solar-driven plumes across the SHT, which would be recorded by new fans. The two hypotheses dependent on internal heat could have also produced new plumes/fans on the SHT, but for those hypotheses and the geologically-short time interval between Voyager 2 and another spacecraft mission to Triton, the new plumes/fans would be expected to be predominantly at the locations of previous eruptions (i.e., locations of Voyager 2 plumes and fans). Therefore, ubiquitous new fan locations on the SHT would be strong evidence for the solar-driven model. The newly formed fans would be identified by comparison with Voyager 2 images. Since Voyager 2 images did not resolve fans outside of the flyby-encounter hemisphere (Section 4), this test applies only for the Voyager 2 flyby-encounter hemisphere of Triton. However, the identification of ≈120 fans in Voyager 2 images (Hansen et al., 1990), interpreted to have an age of less than one Neptune-solar orbit in the context of the solar-driven model (e.g., Kirk et al., 1995), suggests that new fans may be abundant on the Voyager 2 flyby-encounter hemisphere, for the solar-driven hypothesis.

Fans without surrounding volatile-ice (nitrogen, but possibly others such as carbon monoxide, methane, argon, and mixtures) would also constrain the models. The absence of nitrogen-ice around a fan rules out the volatile-ice sheet basal-heat model since it requires thick nitrogen-ice. The solar-driven model also requires nitrogen-ice but since only a thin (a few meters) veneer is strictly necessary (Kirk et al., 1995), it is conceivable that the nitrogen-ice could sublimate without completely obscuring the fan. Thus, the absence of nitrogen-ice rules out the volatile-ice sheet basal-heat model but not necessarily the solar-driven model.

Other distributions of fans, and fans on other regions of Triton, such as the SHT outside of the Voyager 2 flyby-encounter hemisphere and any analogous northern hemisphere terrains, are not as constraining without a better understanding of seasonal volatile transport and fan modification on Triton. As in other tests, there is a neutral result of zero fans, although it is conceivable that some constraints may be derived based on likelihood.

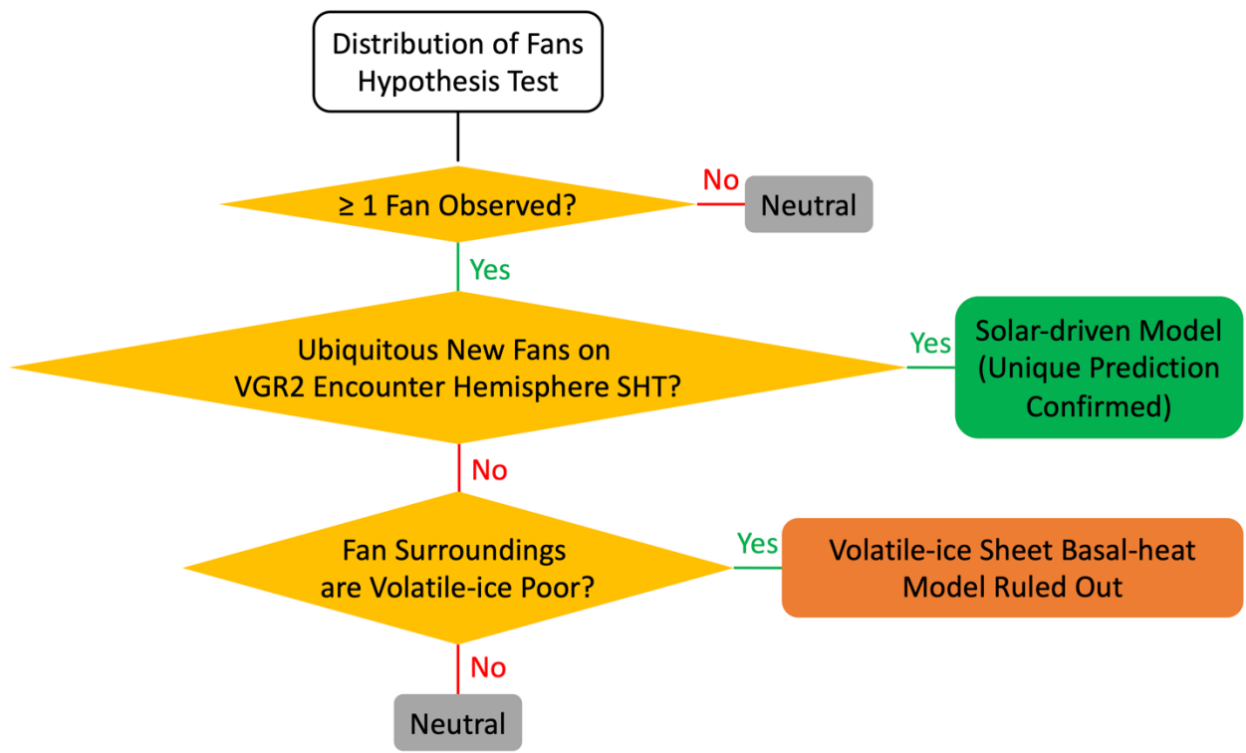

*Figure 8: Distribution of fans hypothesis test. This test assumes the fans are deposits from plumes.*

### 5.5. Anomalous Temperatures Hypothesis Test

A fifth possible test of the hypotheses for Triton's plumes is based on any elevated surface temperatures associated with plumes and/or fans (Figure 9). The solar-driven model predicts solid-state greenhouse heating and sublimation and/or melting of a trapped volume (Brown et al., 1990). The volatile-ice sheet basal-heat hypothesis similarly may or may not include the liquid phase (Section 2.3). It is conceivable that solar insolation and/or internal heat could drive boiling of the liquid but it is more plausible that excess heating would drive further melting and that heat would be transported by the liquid phase, since the hypotheses specifically include eruptions. Thus, the boiling temperature of nitrogen-ice (or other volatile-ice) is a very conservative upper limit for temperatures associated with the solar-driven and volatile-ice sheet basal-heat models. Assuming no isostatic compensation, the thickness of the SHT is ≲ 1 km (Schenk et al., 2021). A ≈1 km thick nitrogen-ice overburden at the surface of Triton results in a pressure of ≈10 bar. At 10 bar, the boiling temperature of nitrogen is 104 K (Air Liquide, 2021). Carbon monoxide and methane would correspond to pressures of ≈10 and ≈5 bar and temperatures of ≈109 and ≈135 K respectively (Air Liquide, 2021). Thus, ≈135 K is a maximum temperature for the solar-driven and volatile-ice sheet basal-heat models. In contrast, the cryovolcanic hypothesis may include liquid water and thus could involve considerably higher temperatures than the other two hypotheses. The temperature of an erupting fissure on Enceladus, for example, was reported to be 197 K (Goguen et al., 2013). Thus, a surface temperature at a plume and/or fan of ≳ 135 K would rule out the solar-driven and volatile-ice sheet basal-heat models.

Note that none of the three hypotheses necessarily predict that elevated surface temperatures will be observed, since the heating may occur far from the surface and/or the surface will cool after the eruption. Thus, only upper limits, not predictions, are used in the anomalous temperatures hypothesis test. Observation of anomalous surface temperatures could be accomplished with a dedicated thermal instrument, as demonstrated with the Cassini CIRS instrument at Enceladus (e.g., Howett et al., 2011), but for some temperatures, measurement may also be possible with a near-infrared spectrometer, as demonstrated at Enceladus with the Cassini VIMS instrument (Goguen et al., 2013).

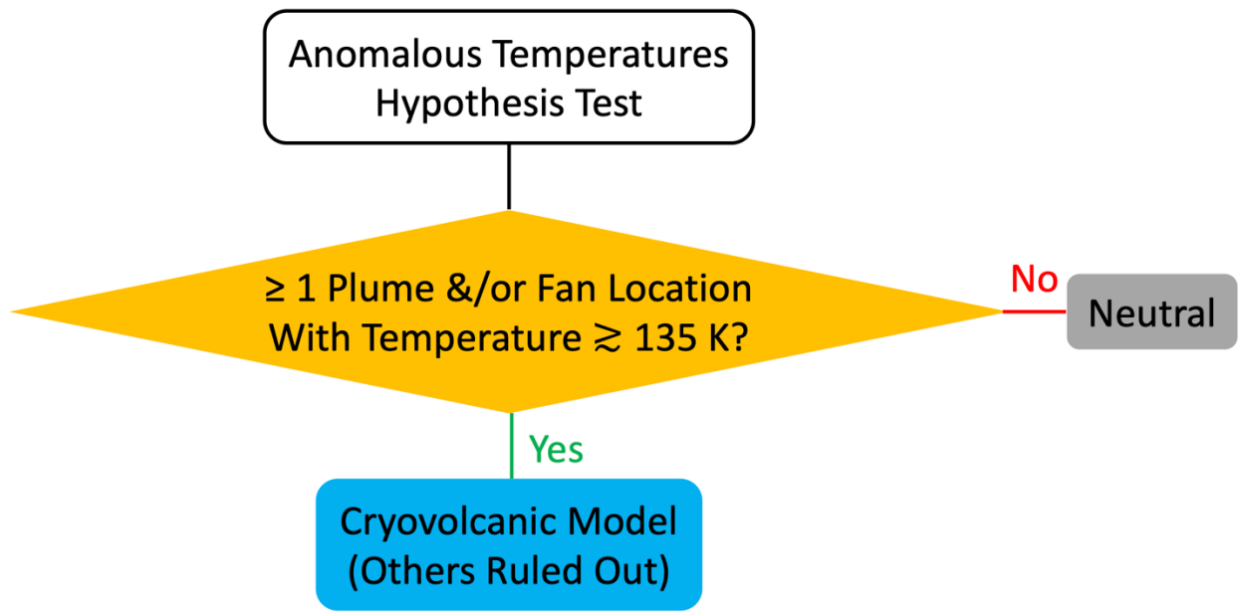

*Figure 9: Anomalous temperatures hypothesis test.*

The five tests of hypotheses for Triton's plumes (Table 2; Figures 5-9) are independent, i.e., each could potentially determine the process responsible for Triton's plumes. They are, however, also complementary and can be combined. The same conclusion from multiple tests would further increase confidence in the result. None of the tests can confidently confirm the volatile-ice sheet basal-heat hypothesis but this hypothesis could be confirmed by process of elimination, where the other two hypotheses are ruled out by separate tests. We also note that it is possible that two, or even all three, processes operate on Triton (although it is unlikely they would result in similar plumes and fans). This possibility could be identified by different results from different tests and, as appropriate, applying the same test separately to different subsets of plumes/fans. The fan composition test, for example, could be applied to each fan independently from all others, although in practice more fans increases confidence in the measurement and conclusion. Additionally, it is possible that none of the published hypotheses operate on Triton and its plumes and fans result from an as yet unconceived process. This possibility could also be identified, by ruling out all three eruption hypotheses, since each hypothesis could be ruled out by at least one test. These five hypothesis tests could also be supplemented by additional geologic context (e.g., topography, morphology, superposition relationships, etc.); for example, structures similar to the tiger stripes at Enceladus could provide further evidence in support of the cryovolcanic hypothesis.

## 6. Subsolar Latitude on Triton in Coming Decades

The inclination of Triton's orbit of Neptune is 157° (retrograde and inclined by 23°) and the pole precesses relatively rapidly (≈700 Earth-years; Jacobson, 2009) relative to the period of Neptune's solar orbit (≈165 Earth-years). In combination with Neptune's solar obliquity of 28°, this results in significant variation of Triton's seasonal excursion of subsolar latitude (effective solar obliquity of Triton) from ≈5-51° (≈28 +/- 23°). Thus, some summer/winter seasons on Triton are dramatically more extreme than others. The subsolar latitude on Triton during the 1989 Voyager 2 encounter was 45° S; it was moving southward and continued to do so until southern summer solstice in 2001 at 50° S, an extreme southern summer. Extreme southern summers where the subsolar latitude was closer to the south pole than 40° S had not previously occurred for approximately half a millennium and will not occur again for even longer (Kirk et al., 1995 and references therein). This century therefore offers a rare opportunity to study the influence of extreme southern summer and extreme northern winter on Triton's seasonal processes. In particular, we argue below that the next few decades are an especially opportune time for studying seasonal impacts on Triton's plumes.

Figure 10A shows the subsolar latitude on Triton from 1980-2100 (the longer-term variation is shown in Soderblom et al. (1990) and Kirk et al. (1995) and references therein). The northern margin of Triton's SHT, which confines all of the plumes and fans (Section 4), varies with longitude from ≈25° S to 5° N (Figure 4A; full longitudinal coverage is shown in McEwen (1990) and in Hillier et al. (1994)). This range, which we consider to be approximate, since the Voyager 2 images have lower spatial resolution outside of the flyby-encounter hemisphere and the margin may have changed with time, is also indicated in Figure 10A. Using 25° S to 5° N as the SHT margin, the subsolar latitude will not be directly over the SHT at some longitudes starting in ≈2029, and will entirely cease to be directly over the SHT by ≈2049. We calculated the fraction of longitudes that are covered by SHT, as a function of latitude, using the map in McEwen (1990). The fractional SHT surface area at the subsolar latitude is shown in Figure 10B. The large change over an approximately two-decade-long period is an important consideration in the context of the solar-driven model for Triton's plumes, as it suggests the next few decades may correspond to a last chance to observe plumes actively erupting from the SHT. The subsolar latitude will not return to the SHT for approximately a century. No analogous northern hemisphere terrains have been observed. If such a terrain exists, it must have been less extensive than the SHT at the time of the Voyager 2 encounter, since Voyager 2 images of the surface extended to ≈45° N.

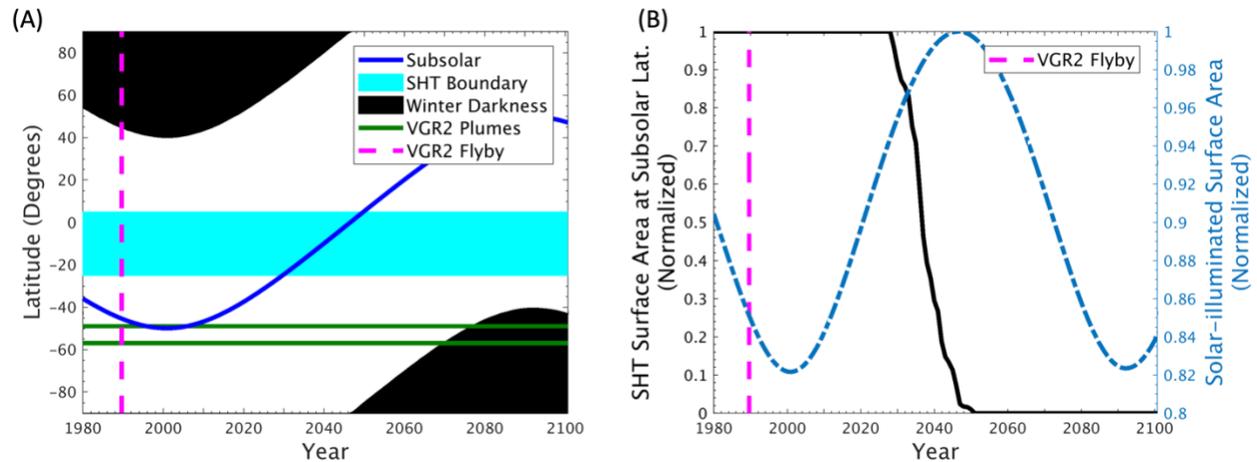

*Figure 10: The changing subsolar latitude on Triton: implications for solar-driven plumes and other seasonal considerations. (A) The subsolar latitude on Triton (blue line). The cyan region indicates the latitude range of the northern boundary of Triton's southern hemisphere terrains (SHT; the boundary varies with longitude) during the Voyager 2 flyby in 1989. The black regions indicate latitudes that do not receive direct solar illumination due to winter darkness (polar night). The green horizontal lines indicate the latitudes of the Mahilani and Hili plumes. (B) The black line indicates the longitudinal fractional area mapped as SHT at the subsolar latitude (which depends on date as shown in (A)). The amount of SHT at the subsolar latitude drops precipitously from ≈2030-2050. The fractional surface area of Triton that receives direct solar illumination is also shown (dashed, blue line and right ordinate axis). The epoch of ≈2025-2050 is an opportune time to explore Triton before the subsolar latitude leaves the SHT and winter darkness begins in the south polar region.*

The changing subsolar latitude also affects which Triton latitudes receive direct solar illumination: continuous winter darkness in the polar regions (polar night) lasts far longer and extends to far more latitudes during Triton's extreme seasons. In addition to seasonal processes, this affects spacecraft observations. Latitudes in winter darkness are shown in Figure 10A and the fraction of Triton's surface that receives direct solar illumination is shown in Figure 10B. Approximately 20% of Triton's surface is in winter darkness during extreme winters. Thus, an area comparable to nearly half that of Canada may not be observable with instruments that depend on solar illumination (however, Neptune-shine increases the observable area, as at Pluto with Charon-shine; Lauer et al., 2021; scattering by atmospheric hazes may also appreciably extend the observable area, also as at Pluto; Schenk et al., 2018). Figure 10B shows that the total illuminated area is currently rising toward 100% of Triton's surface in 2046, after that Triton's south pole will not receive direct solar illumination for nearly a century.

The latitudes of the two confirmed Voyager 2 plumes are also indicated in Figure 10A. These sites of special interest will be in continuous winter darkness toward the end of this century. The minimum incidence angle of solar illumination at these sites (given by the separation between the blue subsolar latitude and green plume-latitude curves) will rise until then. The quality of imaging tends to degrade as incidence angle becomes small or large. In the case of Triton's plumes, however, incidence angles near 90º may enable forward scattering observations that are very favorable. In the absence of direct solar illumination, Neptune-shine may enable lower-flux observation of the sites of the two confirmed plumes, since they are on Triton's Neptune-facing hemisphere.

Taking the above considerations together, Figure 10 suggests that the 2030s and 2040s are an opportune time to explore Triton from a seasonal perspective. This epoch corresponds to (1) a last chance for approximately a century to observe the SHT while it is at the subsolar latitude and hence may be the last occurrence of plumes in the solar-driven model, (2) the total solar-illuminated area is near its maximum and the south pole receives direct sunlight before an almost century-long polar night, and (3) good solar illumination of the Mahilani and Hili plume regions. It is also an opportunity to observe the northern hemisphere as an extreme winter concludes. We consider the above arguments to be good motivation for observing Triton in this period but not an imperative given the scientific caveats to (1) noted earlier and technological solutions (such as radar and thermal imaging) to (2) and (3) as well as the possible Neptune-shine solution to (3) and part of (2).

## 7. Conclusions

The similarity of the latitude of Triton's plumes to the subsolar latitude during the Voyager 2 encounter is not a prediction of the solar-driven hypothesis. The similar latitudes can be consistent with the solar-driven model if the model parameters are tuned; however, it is not strong evidence in favor of this model. The incidence angle dependence of bolometric albedo does not have a significant effect on this conclusion.

The distribution of Triton's fans in areas observed at high spatial resolution by Voyager 2 is controlled by the distribution of its southern hemisphere terrains (SHT). The observation that the fans were generally north of the active plumes, and at latitudes consistent with former plumes that erupted near the subsolar latitude in the decades preceding the Voyager 2 imaging, may be attributed entirely to a broad fan distribution over the SHT and observational bias of the Voyager 2 imaging coverage. The geographic distribution of the fans is strong evidence that the fans (and plumes, assuming the fans are from former plumes) and SHT are related but, with the current uncertainty of SHT properties, including composition and thickness, it is not strong evidence in favor of the solar-driven model.

The solar-driven hypothesis for Triton's plumes was, for three decades, generally considered the leading hypothesis, however, the above two conclusions contest two prevalent arguments in favor of that hypothesis. A third argument in favor of the solar-driven hypothesis, that the high-albedo SHT is nitrogen-rich, has also been questioned based on astronomical observations (Grundy et al., 2010; Holler et al., 2016). The greater longitudinal variability of spectral features of volatile-ices (nitrogen, carbon monoxide, and methane) than that of non-volatile-ices (water and carbon dioxide) is interpreted as evidence that the southern hemisphere is dominated by non-volatile-ices. The astronomical observations, however, were not contemporaneous with the Voyager 2 observations. We conclude that all three viable hypotheses for Triton's plumes, the solar-driven, cryovolcanic, and volatile-ice sheet basal-heat eruption models, warrant further consideration.

If Triton's plumes are explosive cryovolcanic eruptions then either (1) the fans are not expected to be from such plumes and/or (2) the SHT is not expected to be enriched in volatile-ice. Otherwise, the fans would not be expected to be broadly distributed but also strongly confined to the SHT, as was the case for all of the > 100 fans observed by Voyager 2.

The solar-driven, cryovolcanic, and volatile-ice sheet basal-heat hypotheses for Triton's plumes can likely be confidently distinguished with another single-flyby mission. Five tests of the predictions of the models that could be implemented with spacecraft remote sensing are: SHT composition and thickness, fan composition, active plume distribution, fan distribution, and anomalous temperatures. The five tests are independent, but complementary, which increases resiliency.

From a seasonal perspective, the 2030s and 2040s are an opportune time to explore Triton. This epoch corresponds to a last chance for approximately a century to observe the SHT while it is at the subsolar latitude and hence may be the last occurrence of plumes in the solar-driven model and also corresponds to favorable solar-illumination for global observations.

**Acknowledgements:** The Trident mission concept inspired this research; we thank the entire concept team. We thank Randy Persinger at The Aerospace Corporation and Lin Midkiff at Raytheon Technologies for numerous conversations that improved the presentation of this research. We thank Lizbeth de la Torre and Lisa Poje at the Jet Propulsion Laboratory for creating Figure 2. Part of this research was carried out at the Jet Propulsion Laboratory, California Institute of Technology, under a contract with the National Aeronautics and Space Administration (80NM0018D0004). Free and open source QGIS software and the Jet Propulsion Laboratory's HORIZONS On-Line Ephemeris System were used in this research. The service of two anonymous reviewers and the editor is gratefully acknowledged.

**Bibliography**

Air Liquide 2021. Gas Encyclopedia by Air Liquide. https://encyclopedia.airliquide.com/

Agnor, C. B., Hamilton, D. P., May 2006. Neptune's capture of its moon Triton in a binary-planet gravitational encounter. Nature 441 (7090), 192-194.

Brown, R. H., Kirk, R. L., Jan. 1994. Coupling of volatile transport and internal heat flow on Triton. Journal of Geophysical Research 99 (E1), 1965-1982.

Brown, R. H., Kirk, R. L., Johnson, T. V., Soderblom, L. A., Oct. 1990. Energy Sources for Triton's Geyser-Like Plumes. Science 250 (4979), 431-435.

Buratti, B. J., Bauer, J. M., Hicks, M. D., Hillier, J. K., Verbiscer, A., Hammel, H., Schmidt, B., Cobb, B., Herbert, B., Garsky, M., Ward, J., Foust, J., Apr. 2011. Photometry of Triton 1992-2004: Surface volatile transport and discovery of a remarkable opposition surge. Icarus 212 (2), 835-846.

Buratti, B. J., Hicks, M. D., Dalba, P. A., Chu, D., O'Neill, A., Hillier, J. K., Masiero, J., Banholzer, S., Rhoades, H., May 2015. Photometry of Pluto 2008-2014: Evidence of Ongoing Seasonal Volatile Transport and Activity. Astrophysical Journal Letters 804 (1), L6.

Croft, S. K., Kargel, J. S., Kirk, R. L., Moore, J. M., Schenk, P. M., Strom, R. G., Jan. 1995. The geology of Triton. In: Neptune and Triton. pp. 879-947.

Duxbury, N. S., Brown, R. H., Jan. 1997. The Role of an Internal Heat Source for the Eruptive Plumes on Triton. Icarus 125 (1), 83-93.

Fray, N., Schmitt, B., Dec. 2009. Sublimation of ices of astrophysical interest: A bibliographic review. Planetary and Space Science 57 (14-15), 2053-2080.

Geissler, P., 2015. Chapter 44 - Cryovolcanism in the outer solar system. In: Sigurdsson, H. (Ed.), The Encyclopedia of Volcanoes (Second Edition). Academic Press, Amsterdam, pp. 763-776.

Goguen, J. D., Buratti, B. J., Brown, R. H., Clark, R. N., Nicholson, P. D., Hedman, M. M., Howell, R. R., Sotin, C., Cruikshank, D. P., Baines, K. H., Lawrence, K. J., Spencer, J. R., Blackburn, D. G., Sep. 2013. The temperature and width of an active fissure on Enceladus measured with Cassini VIMS during the 14 April 2012 South Pole flyover. Icarus 226 (1), 1128-1137.

Grundy, W. M., Binzel, R. P., Buratti, B. J., Cook, J. C., Cruikshank, D. P., Dalle Ore, C. M., Earle, A. M., Ennico, K., Howett, C. J. A., Lunsford, A. W., Olkin, C. B., Parker, A. H., Philippe, S., Protopapa, S., Quirico, E., Reuter, D. C., Schmitt, B., Singer, K. N., Verbiscer, A. J., Beyer, R. A., Buie, M. W., Cheng, A. F., Jennings, D. E., Linscott, I. R., Parker, J. W., Schenk, P. M., Spencer, J. R., Stansberry, J. A., Stern, S. A., Throop, H. B., Tsang, C. C. C., Weaver, H. A., Weigle, G. E., Young, L. A., Mar. 2016. Surface compositions across Pluto and Charon. Science 351 (6279), aad9189.

Grundy, W. M., Young, L. A., Stansberry, J. A., Buie, M. W., Olkin, C. B., Young, E. F., Feb. 2010. Near-infrared spectral monitoring of Triton with IRTF/SpeX II: Spatial distribution and evolution of ices. Icarus 205 (2), 594-604.

Hansen, C. J., Byrne, S., Portyankina, G., Bourke, M., Dundas, C., McEwen, A., Mellon, M., Pommerol, A., Thomas, N., Aug. 2013. Observations of the northern seasonal polar cap on Mars: I. Spring sublimation activity and processes. Icarus 225 (2), 881-897.

Hansen, C. J., Castillo-Rogez, J., Grundy, W., Hofgartner, J. D., Martin, E. S., Mitchell, K., Nimmo, F., Nordheim, T. A., Paty, C., Quick, L. C., Roberts, J. H., Runyon, K., Schenk, P., Stern, A., Umurhan, O., Aug. 2021. Triton: Fascinating Moon, Likely Ocean World, Compelling Destination! The Planetary Science Journal 2 (4), 137.

Hansen, C. J., McEwen, A. S., Ingersoll, A. P., Terrile, R. J., Oct. 1990. Surface and Airborne Evidence for Plumes and Winds on Triton. Science 250 (4979), 421-424.

Hansen, C. J., Thomas, N., Portyankina, G., McEwen, A., Becker, T., Byrne, S., Herkenhoff, K., Kieffer, H., Mellon, M., Jan. 2010. HiRISE observations of gas sublimation-driven activity in Mars' southern polar regions: I. Erosion of the surface. Icarus 205 (1), 283-295.

Hendrix, A. R., Hurford, T. A., Barge, L. M., Bland, M. T., Bowman, J. S., Brinckerhoff, W., Buratti, B. J., Cable, M. L., Castillo-Rogez, J., Collins, G. C., Diniega, S., German, C. R., Hayes, A. G., Hoehler, T., Hosseini, S., Howett, C. J. A., McEwen, A. S., Neish, C. D., Neveu, M., Nordheim, T. A., Patterson, G. W., Patthoff, D. A., Phillips, C., Rhoden, A., Schmidt, B. E., Singer, K. N., Soderblom, J. M., Vance, S. D., Jan. 2019. The NASA Roadmap to Ocean Worlds. Astrobiology 19 (1), 1-27.

Hillier, J., Veverka, J., Helfenstein, P., Lee, P., Jun. 1994. Photometric Diversity of Terrains on Triton. Icarus 109 (2), 296-312.

Hofgartner, J. D., Buratti, B. J., Devins, S. L., Beyer, R. A., Schenk, P., Stern, S. A., Weaver, H. A., Olkin, C. B., Cheng, A., Ennico, K., Lauer, T. R., McKinnon, W. B., Spencer, J., Young, L. A., New Horizons Science Team, Mar. 2018. A search for temporal changes on Pluto and Charon. Icarus 302, 273-284.

Hofgartner, J. D., Buratti, B. J., Hayne, P. O., Young, L. A., Dec. 2019. Ongoing resurfacing of KBO Eris by volatile transport in local, collisional, sublimation atmosphere regime. Icarus 334, 52-61.

Holler, B. J., Young, L. A., Grundy, W. M., Olkin, C. B., Mar. 2016. On the surface composition of Triton's southern latitudes. Icarus 267, 255-266.

Howett, C. J. A., Spencer, J. R., Pearl, J., Segura, M., Mar. 2011. High heat flow from Enceladus' south polar region measured using 10-600 cm$^{-1}$ Cassini/CIRS data. Journal of Geophysical Research (Planets) 116 (E3), E03003.

Ingersoll, A. P., Mar. 1990. Dynamics of Triton's atmosphere. Nature 344 (6264), 315-317.

Ingersoll, A. P., Tryka, K. A., Oct. 1990. Triton's Plumes: The Dust Devil Hypothesis. Science 250 (4979), 435-437.

Jacobson, R. A., May 2009. The Orbits of the Neptunian Satellites and the Orientation of the Pole of Neptune. Astronomical Journal 137 (5), 4322-4329.

Kieffer, H. H., Christensen, P. R., Titus, T. N., Aug. 2006. CO2 jets formed by sublimation beneath translucent slab ice in Mars' seasonal south polar ice cap. Nature 442 (7104), 793-796.

Kirk, R. L., Brown, R. H., Soderblom, L. A., Oct. 1990. Subsurface Energy Storage and Transport for Solar-Powered Geysers on Triton. Science 250 (4979), 424-429.

Kirk, R. L., Soderblom, L. A., Brown, R. H., Kieffer, S. W., Kargel, J. S., Jan. 1995. Triton's plumes: discovery, characteristics, and models. In: Neptune and Triton. pp. 949-989.

Kite, E. S., Rubin, A. M., Apr. 2016. Sustained eruptions on Enceladus explained by turbulent dissipation in tiger stripes. Proceedings of the National Academy of Science 113 (15), 3972-3975.

Lauer, T. R., Spencer, J. R., Bertrand, T., Beyer, R. A., Runyon, K. D., L White, O., Young, L. A., Ennico, K., McKinnon, W. B., Moore, J. M., Olkin, C. B., Stern, S. A., Weaver, H. A., Oct. 2021. The Dark Side of Pluto. The Planetary Science Journal 2 (5), 214.

McEwen, A. S., Sep. 1990. GLOBAL COLOR AND ALBEDO VARIATIONS ON TRITON. Geophysical Research Letters 17 (10), 1765-1768.

McKinnon, W. B., Sep. 1984. On the origin of Triton and Pluto. Nature 311 (5984), 355-358.

McKinnon, W. B., Nimmo, F., Wong, T., Schenk, P. M., White, O. L., Roberts, J. H., Moore, J. M., Spencer, J. R., Howard, A. D., Umurhan, O. M., Stern, S. A., Weaver, H. A., Olkin, C. B., Young, L. A., Smith, K. E., Beyer, R., Buie, M., Buratti, B., Cheng, A., Cruikshank, D., Dalle Ore, C., Gladstone, R., Grundy, W., Lauer, T., Linscott, I., Parker, J., Porter, S., Reitsema, H., Reuter, D., Robbins, S., Showalter, M., Singer, K., Strobel, D., Summers, M., Tyler, L., Banks, M., Barnouin, O., Bray, V., Carcich, B., Chaikin, A., Chavez, C., Conrad, C., Hamilton, D., Howett, C., Hofgartner, J., Kammer, J., Lisse, C., Marcotte, A., Parker, A., Retherford, K., Saina, M., Runyon, K., Schindhelm, E., Stansberry, J., Steffl, A., Stryk, T., Throop, H., Tsang, C., Verbiscer, A., Winters, H., Zangari, A., New Horizons Geology, Geophysics, and Imaging Theme Team, Jun. 2016. Convection in a volatile nitrogen-ice-rich layer drives Pluto's geological vigour. Nature 534 (7605), 82-85.

Newman, S. F., Buratti, B. J., Brown, R. H., Jaumann, R., Bauer, J., Momary, T., Feb. 2008. Photometric and spectral analysis of the distribution of crystalline and amorphous ices on Enceladus as seen by Cassini. Icarus 193 (2), 397-406.

Nimmo, F., Spencer, J. R., Jan. 2015. Powering Triton's recent geological activity by obliquity tides: Implications for Pluto geology. Icarus 246, 2-10.

Porco, C. C., Helfenstein, P., Thomas, P. C., Ingersoll, A. P., Wisdom, J., West, R., Neukum, G., Denk, T., Wagner, R., Roatsch, T., Kieffer, S., Turtle, E., McEwen, A., Johnson, T. V., Rathbun, J., Veverka, J., Wilson, D., Perry, J., Spitale, J., Brahic, A., Burns, J. A., Del Genio, A. D., Dones, L., Murray, C. D., Squyres, S., Mar. 2006. Cassini Observes the Active South Pole of Enceladus. Science 311 (5766), 1393-1401.

Schenk, P., Beddingfield, C., Bertrand, T., Bierson, C., Beyer, R., Bray, V., Cruikshank, D., Grundy, W., Hansen, C., Hofgartner, J., Martin, E., McKinnon, W., Moore, J., Robbins, S., Runyon, K., Singer, K., Spencer, J., Stern, S., Stryk, T., Sep. 2021. Triton: Topography and Geology of a Probable Ocean World with Comparison to Pluto and Charon. Remote Sensing 13 (17), 3476.

Schenk, P. M., Beyer, R. A., McKinnon, W. B., Moore, J. M., Spencer, J. R., White, O. L., Singer, K., Nimmo, F., Thomason, C., Lauer, T. R., Robbins, S., Umurhan, O. M., Grundy, W. M., Stern, S. A., Weaver, H. A., Young, L. A., Smith, K. E., Olkin, C., New Horizons Geology, Geophysics Investigation Team, Nov. 2018. Basins, fractures and volcanoes: Global cartography and topography of Pluto from New Horizons. Icarus 314, 400-433.

Schenk, P. M., Zahnle, K., Dec. 2007. On the negligible surface age of Triton. Icarus 192 (1), 135-149.

Soderblom, L. A., Kieffer, S. W., Becker, T. L., Brown, R. H., Cook, A. F., I., Hansen, C. J., Johnson, T. V., Kirk, R. L., Shoemaker, E. M., Oct. 1990. Triton's Geyser-Like Plumes: Discovery and Basic Characterization. Science 250 (4979), 410-415.

Squyres, S. W., Veverka, J., Apr. 1982. Variation of albedo with solar incidence angle on planetary surfaces. Icarus 50 (1), 115-122.

Stone, E. C., Miner, E. D., Dec. 1989. The Voyager 2 Encounter with the Neptunian System. Science 246 (4936), 1417-1421.

Tegler, S. C., Grundy, W. M., Olkin, C. B., Young, L. A., Romanishin, W., Cornelison, D. M., Khodadadkouchaki, R., May 2012. Ice Mineralogy across and into the Surfaces of Pluto, Triton, and Eris. The Astrophysical Journal 751 (1), 76.